\documentclass{aa} 
\usepackage[colorlinks = true,
            linkcolor = blue,
            urlcolor  = blue,
            citecolor = blue,
            anchorcolor = blue]{hyperref}
\usepackage{natbib}
\usepackage[ ]{caption}
\bibpunct{(}{)}{;}{a}{}{,} 
\usepackage{graphicx}
\usepackage{txfonts}
\begin{document}

   \title{ISPY - NaCo Imaging Survey for Planets around Young stars\thanks{ESO program IDs 097.C-0206 and 1101.C-0092}}

   \subtitle{Discovery of an M dwarf in the gap between HD\,193571 and its debris ring}

   \author{Arianna Musso Barcucci\inst{1}
          \and Ralf Launhardt\inst{1}
          \and Grant M. Kennedy\inst{2}
          \and Henning Avenhaus\inst{3}
          \and Stefan S. Brems\inst{4}
          \and \\Roy van Boekel\inst{1}
          \and Faustine Cantalloube\inst{1}
          \and Anthony Cheetham\inst{1,5}
          \and Gabriele Cugno\inst{6}
          \and Julien Girard\inst{7}
          \and Nicol\'as Godoy\inst{8,9}
          \and Thomas K. Henning\inst{1}
          \and Stanimir Metchev\inst{10}
          \and Andr\'e M\"uller\inst{1}
          \and Johan Olofsson\inst{8,9}
          \and Francesco Pepe\inst{5}
          \and Sascha P. Quanz\inst{6}
          \and Andreas Quirrenbach\inst{4}
          \and Sabine Reffert\inst{4}
          \and Emily L. Rickman\inst{5}
          \and Matthias Samland\inst{1}
          \and Damien Segransan\inst{5}
          }

   \institute{Max Planck Institute for Astronomy (MPIA),
              K\"onigstuhl 17, 69117 Heidelberg, Germany\\
              \email{musso@mpia.de}
        \and
        Department of Physics, University of Warwick, Gibbet Hill Road, Coventry CV4 7AL, UK
        \and
        Lakeside Labs, Lakeside Park B04b, A-9020 Klagenfurt, Austria
        \and
        Landessternwarte, Zentrum f\"ur Astronomie der Universit\"at Heidelberg, K\"onigstuhl 12, 69117 Heidelberg, Germany
        \and
        Observatoire Astronomique de l'Universit\'e de Gen\`eve, 51 Ch. des Maillettes, 1290 Versoix, Switzerland
        \and
        ETH Z\"urich, Institute for Particle Physics and Astrophysics, Wolfgang-Pauli-Str. 27, 8093 Z\"urich, Switzerland
        \and
        European Southern Observatory (ESO), Karl-Schwazschild-Str. 2, 85748 Garching, Germany
        \and
        Instituto de Física y Astronomía, Facultad de Ciencias, Universidad de Valparaíso, Av. Gran Bretaña 1111, Playa Ancha, Valparaíso, Chile - \email{johan.olofsson@uv.cl}
        \and
        N\'ucleo Milenio Formaci\'on Planetaria - NPF, Universidad de Valparaíso, Av. Gran Bretaña 1111, Playa Ancha, Valparaíso, Chile
        \and
        The University of Western Ontario, Dept. of Physics and Astronomy, 1151 Richmond Avenue, London, ON N6A 3K7, Canada
      }

   \date{Received: 1 February 2019 / Accepted: 20 May 2019}

  \abstract
   {The interaction between low-mass companions and the debris discs they reside in is still not fully understood.
    A debris disc can evolve due to self-stirring, a process in which planetesimals can excite their neighbours to the point of destructive collisions. In addition, the presence of a companion could further stir the disc (companion-stirring).
Additional information is necessary to understand this fundamental step in the formation and evolution of a planetary system, and at the moment of writing only a handful of systems are known where  a companion and a debris disc have both been detected and studied at the same time.}{Our primary goal is to augment the sample of these systems and to understand the relative importance between self-stirring and companion-stirring.}
{In the course of the VLT/NaCo  Imaging Survey for Planets around Young stars (ISPY), we observed HD\,193571, an A0 debris disc hosting star at a distance of $68$ pc with an age between $\sim$60 and 170 Myr.
We obtained two sets of observations in $L$' band and a third epoch in $H$ band using the GPI instrument at Gemini-South.}{A companion was detected in all three epochs at a projected separation of $\sim$11 au ($\sim$0.17"), and co-motion was confirmed through proper motion analysis. Given the inferred disc size of $120$ au, the companion appears to reside within the gap between the host star and the disc. Comparison between the $L$' and $H$ band magnitude and evolutionary tracks suggests a mass of $\sim$$0.31-0.39\,M_{\odot}$.}{We  discovered a previously unknown M-dwarf companion around HD\,193571, making it the third low-mass stellar object discovered within a debris disc. A comparison to self- and companion-stirring models suggests that the companion is likely responsible for the stirring of the disc.}
\titlerunning{Discovery of an M-dwarf around HD\,193571}
   \maketitle
%

\section{Introduction}
Circumstellar discs are the natural by-products of the protostellar accretion process and they are the birthplaces of planetary systems.
They evolve with time and undergo a series of processes. After the material that was in the original protoplanetary disc has been dissipated 
\mbox{\citep[usually within $\sim10$ Myr, see][]{ErcolanoPascucci2017}}, a new generation of dust is created and continuously replenished via planetesimal collisions, forming a second generation debris disc (DD). These destructive encounters are triggered when the planetesimals are dynamically excited such that their relative velocities increase above a critical value (low-velocity collisions can happen in non-excited DDs as well, but they  produce a different and recognisable emission spectrum, see  \citealt{Heng_Tremaine_2010}). Three possible stirring processes have been proposed so far that could induce such an excitation in the disc: stellar encounters, self-stirring and companion-stirring. Of these three, the first scenario is the least likely one to be observed, since close stellar encounters are rare (particularly among field stars) and the disc brightness resulting from dust production drops too quickly to be detectable \mbox{\citep{KenyonBromley2002}}.\\ In the self-stirring scenario \mbox{\citep{KenyonBromley2008}}, planetesimals with low relative velocities form increasingly large  bodies that in return dynamically excite smaller neighbours above the critical threshold for planetesimal destruction. The planetesimal growth  scales with orbital period, resulting in an inside-out collisional cascade. Since a maximum growth speed is set by the host star and disc parameters, at any given time there is a maximum disc size that can be explained by self-stirring.\\ In the companion-stirring case \mbox{\citep{MustillWyatt2009}}, 
the planetesimals are excited by  the companion's secular perturbations, and the maximum disc size at a given time is a function of the physical properties of both the central star and the companion.\\
The optimal scenario to investigate these processes is therefore the one in which  the disc and the companion(s) are observed and characterised at the same time.
At the moment of writing, only a handful of such systems have been found:  HR\,8799  is one of the most  extensively studied \mbox{\citep{hr8799_marois_2008}}, alongside HD\,95086 \mbox{\citep{hd95086_rameau2013discovery}} and $\beta$\,Pic \mbox{\citep{Lagrange_2010}}.
In addition, only two systems are currently known  where the companion is in the stellar mass regime: HR\,2562 \mbox{\citep{Konopacky2016}} and HD\,206893 \mbox{\citep{Milli2017_HD206893}}.\\
The limited number of systems suitable to investigate the companion-disc interaction 
does not allow us to fully comprehend this phenomenon, and therefore augmenting this sample is our primary goal.\\
Detecting and characterising giant planets (GPs) around DD host stars is one of the scientific goals of the Imaging Survey for Planets around Young Stars (ISPY, Launhardt et al., in prep.), currently underway at the Very Large Telescope (VLT). It makes use of the NaCo instrument \mbox{\citep{Lenzen2003_naco,Rousset2003_naco}} in $L'$ band, and the Angular Differential Imaging  \mbox{\citep[ADI,][]{adi_marois2006}}.\\
\section{HD\,193571}
Within the NaCo-ISPY survey, we observed HD\,193571 (HR\,7779, GJ\,969, $\kappa\,$01 Sgr), an A0V field star at a distance of 68.45\,pc {\citep{Gaia_2018}}, which is part of a wide-separation ($>$40") three-component system\mbox{\footnote{The B and C components were observed in   2000 and 1999, and have a distance of $39.30"$ and $56.80"$, with a P.A. of $312^{\circ}$ and $283^{\circ}$, respectively.} }\mbox{\citep[WDS Catalogue, see][]{Mason_2001_2014}}.\\
The age of this target is uncertain: \mbox{\citet{DavidHillenbrand2015}} derived stellar parameters for more than 3000 nearby early-type (BAF) field stars, and compared them with stellar isochrones. They computed final ages and masses with both a Bayesian inference approach and classical isochrone interpolation, obtaining 161 Myr and 66 Myr, respectively.
They presented criteria to decide between the two values, but for 
HD\,193571 it is unclear which age or mass estimate should be preferred.
Throughout this study we  use a primary mass of $M=2.2\pm0.1\,M_{\odot}$, which encompasses both the Bayesian inferred mass and the mass derived via interpolation. \\
The age estimates for HD\,193571 are presented in \autoref{table: basic}, together with the main stellar properties.

\begin{table}
\caption{Fundamental stellar parameters and properties for HD\,193571.}             
\label{table: basic}      
\centering                          
\begin{tabular}{l l c} 
\hline \hline
Parameter  &   Value  &   Ref.\\
\hline
RA [hh:mm:ss] & 20:22:27.50 & 5\\
DEC [dd:mm:ss] & -42:02:58.43 & 5\\
Parallax [mas] & $14.61 \pm 0.17$ & 1\\
Distance [pc] & $68.45 \pm 0.82$ & 1\\
Proper motion [mas/yr] & $\muup_{\alpha}\times cos\delta = 41.31 \pm 0.22$ & 1\\
                            &  $\muup_{\delta} = -83.74 \pm 0.19$ & 1\\
Sp. Type   & A0V &  6\\
T$_\mathrm{eff}$ [K]    & $9740\pm100$ & 3\\
Mass ${[}M_{\odot}{]}$ & $2.2\pm0.1$ & 2\\
Radius ${[}R_{\odot}{]}$ & $1.85\pm0.1$  & 3\\
$v\,sin\,i$ [km/s] & $71$ & 2\\
L ${[}L_{\odot}{]}$ & $27.7\pm1$  & 3\\
$f=L_\mathrm{disc}/L_{\star}$ & $2.3\times10^{-5}\pm1\times10^{-6}$ & 3\\
Bayesian Age [Myr]  & $161^{+247}_{-35}$ & 2\\
Interp. Age [Myr]  & $66$ & 2\\
$m_\mathrm{L'}$ [mag]  & $5.614\pm0.030$ & 4\\
$m_\mathrm{H}$ [mag]  & $5.609\pm0.030$ & 4\\

\hline                               
\end{tabular}
\bigskip
\caption*{\textbf{References:} (1) \citet{Gaia_2016,Gaia_2018}; (2) \citet{DavidHillenbrand2015}; (3) this work (see Section 2); (4) apparent magnitude of the host star in the $L'$ band, derived from SED fitting (see Section 2.1) and correcting for the NaCo $L'$ band transmission curve; (5) value taken from the online Simbad catalogue; (6) \citet{Chen_2014}.}
\end{table}

\begin{figure*}[t!]
        \centering
                \includegraphics[width=\hsize]{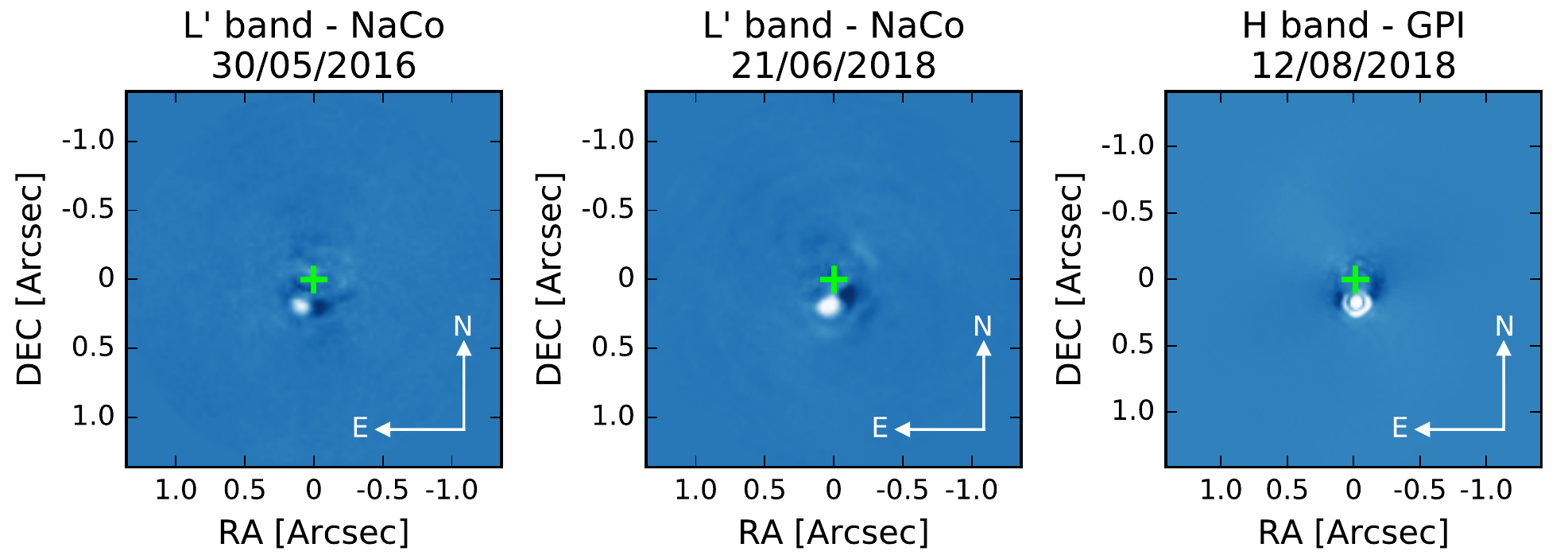}
                \caption{Classically ADI reduced images for the two NaCo datasets (left and centre) and for the GPI dataset (right). The images are oriented with north up and east left, and the green cross indicates the position of the central star. The companion is clearly visible close to the centre in all three datasets. The images are normalised and the colour map was chosen for a better visualisation of the data.}
        \label{Fig: detection}\
\end{figure*}

\begin{figure}[t!]
        \centering
                \includegraphics[width=\hsize]{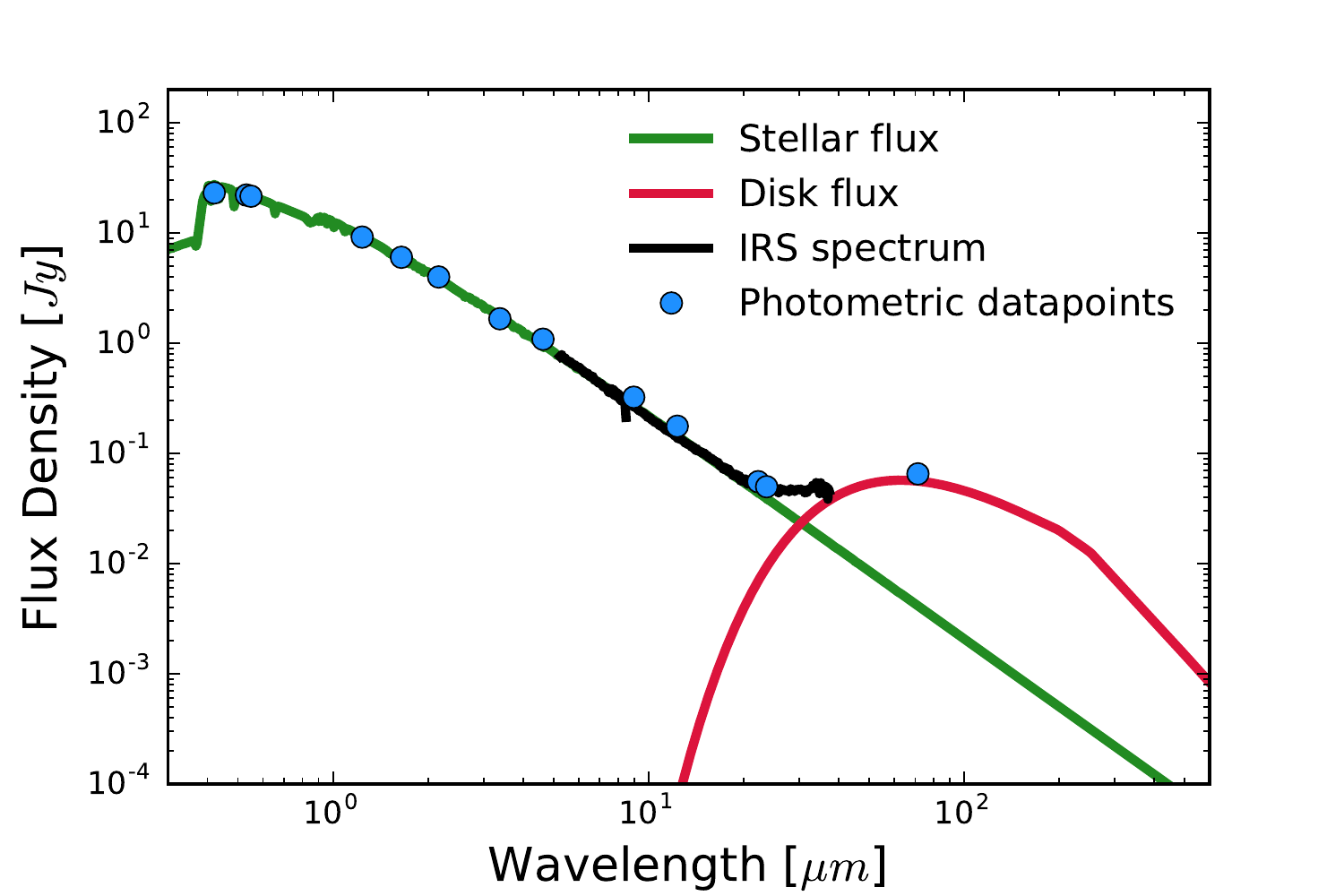}
                \caption{Flux density distribution of HD\,193571, showing the photometric datapoints found in the literature (in blue) and the IRS spectrum (in black), together with the fitted stellar (green) and disc (red) fluxes.}
        \label{Fig: SED}\
\end{figure}

HD\,193571 is known to harbour a debris disc, inferred from its infrared excess \mbox{($f=L_\mathrm{disc}/L_{\star}=2.3\times 10^{-5}$)}.
We fit its spectral energy distribution (SED) to derive the stellar luminosity and effective temperature, and the debris belt radius. We fit
simultaneously a stellar atmosphere \citep[PHOENIX;][]{Husser_2013} plus a single black-body (BB) model to the observed photometry and the Spitzer IRS spectrum. The photometry includes a wide range of filters and wavelengths, from: “Heritage” Stromgren and UBV (\citealt{Paunzen_2015}), 2MASS (\citealt{2MASS_catalogue_2006}), Hipparcos/Tycho-2 (\citealt{Esa_1997}), AKARI (\citealt{Ishihara_2010}), WISE (\citealt{Wright_2010}), and Spitzer (\citealt{Chen_2014}). The fitting method uses synthetic photometry of grids of models, and finds the best-fitting model with the MultiNest code \citep{Feroz_2009}. The SED of HD\,193571 is best fit by an A0 stellar model plus a one-temperature BB model locating the dust at a distance of $R_\mathrm{BB}=62\pm4$ au, with a temperature of $81\pm3$ K. The best fit is shown in \autoref{Fig: SED}.\\
The BB radius of the dust disc is given by \citep{PawellekKrivov2015}
\begin{equation*}
R_{\rm BB} = \Bigg( \frac{278\,K}{T_{\rm
dust}}\Bigg)^2\,\Bigg(\frac{L}{L_{\odot}}\Bigg)^{1/2}
\end{equation*}
An estimate of the `true' disc radius, $R_{\rm disc}$, is then obtained
by applying a stellar luminosity-dependent
correction factor, $\Gamma$, which accounts for the radiation pressure
blowout grain size
\begin{equation*}
\Gamma = a\,( L_{\ast}/{\rm L_{\odot}})^b
\end{equation*}
\citep{PawellekKrivov2015}, using the new coefficients given in
\citet{Pawellek2016thesis}:
$a=7.0$\ and $b=-0.39$.
After applying this correction, the estimated disc size for HD\,193571 is $120\pm15$ au.
The disc has never been imaged in scattered light, and additional SPHERE/IRDIS observations were inconclusive in this respect (see Appendix A).\\

We used the fitted stellar spectrum to derive the stellar $H$ and $L'$ magnitudes (reported in \autoref{table: basic}), integrating over the NaCo $H$- and $L'$-band filters. We used zero points of $1.139\times10^{-10}\,\mathrm{erg}/\mathrm{cm}^{2}/\mathrm{s}/\AA$ and $5.151\times10^{-12}\,\mathrm{erg}/\mathrm{cm}^{2}/\mathrm{s}/\AA$, respectively\footnote{\url{http://svo2.cab.inta-csic.es/svo/theory/fps3/index.php?mode=browse&gname=Paranal&gname2=NACO}}.

\section{Observations and data reduction}
HD\,193571 was observed at two different epochs with NaCo at the Very Large Telescope (VLT), and an additional third epoch was obtained with the Gemini Planet Imager \mbox{\citep[GPI,][]{Macintosh_2014_GPI}} through the Fast Turnaround observing mode (Program ID: GS-2018A-FT-111).
\subsection{$VLT/NaCo$}   
Coronagraphic ADI observations of HD\,193571 were obtained in May 2016 and June 2018 in L$'$ band (see \autoref{table: observations}), making use of the Annular Groove Phase Mask  \mbox{\citep[AGPM,][]{Mawet2013_AGPM}} vector vortex coronagraph to suppress as much as possible the diffraction pattern from the host star. We used cube-mode, saving 100 frames per cube.
The observations were interlaced with frequent sky observations for background subtraction (every $\sim$8 minutes) and bracketed with non-coronagraphic flux measurements to create an unsaturated PSF reference.
The data was reduced with the ISPY end-to-end modular reduction pipeline GRAPHIC \mbox{\citep{graphic}}. The main reduction steps comprise background subtraction, flat field correction, bad pixel cleaning, and centring. Each cosmetically 
reduced cube is then median combined. For a more detailed explanation on how the data reduction is performed we refer to the ISPY overview paper (Launhardt et al., in prep). The observations are summarised in \autoref{table: observations}.\\

\begin{table}
\caption{VLT/NaCo summary of observations}             
\label{table: observations}      
\centering                          
\begin{tabular}{l | c c} 
\hline \hline
Parameter & Epoch 1 & Epoch 2 \\
\hline
Obs.  &  30/05/2016 & 21/06/2018 \\
Prog. ID  &   097.C-0206 & 1101.C-0092 \\
$\#$cubes & 91 & 196 \\
Tot. P.A. & $78^{\circ}$ & $84^{\circ}$ \\
DIT Obs.$^{a}$ [s] & 0.35 & 0.35 \\
DIT Flux$^{b}$ [s] & 0.07 & 0.07 \\
DIMM$^{c}$ & $\sim$$1\farcs0$ & $\sim$$1\farcs1$ \\
Tot. time$^{d}$ [m] & 53 & 114\\
Sky time$^{e}$ [m] & 4.1 & 9.3\\
\hline 
\end{tabular}
\bigskip
\caption*{$^{a}$ Detector Integration Time for the observations, chosen to avoid saturation outside $\sim0\farcs1$. $^{b}$ Detector Integration Time for the non-coronographic flux measurements. $^{c}$ Mean DIMM seeing during the observations. $^{d}$ Total on-source integration time, in minutes.\\$^{e}$ Total on-sky time, in minutes: 7 sky visits for the 2016 dataset and 16 sky visits for the 2018 dataset.
}
\end{table}

\subsection{Gemini/GPI}  
HD\,193571 was observed in the $H$ band with GPI in coronagraphic ADI mode on  12  August 2018, obtaining 76 frames and achieving a total field rotation of $88$ degrees. The integration time for each exposure was 60 seconds.\\
The photometry of GPI data can be calibrated using the satellite spots, which are four reference spots created by diffraction of the central star light from a square grid superimposed on the pupil plane (\citealt{Wang_2014}). They can be used to extract the photometry and spectroscopy of the central star. During the observations there was a misalignment of the grid that produces the satellite spots, resulting in a diffraction spike above  two of the four satellite spots, thus rendering them unusable for photometric calibration. Therefore, in the following analysis, when referring to the satellite spots, we  only refer  to the two unbiased ones.\\
The data were reduced making use of the publicly available GPI Data Pipeline \mbox{\citep{Maire_2010}}, with the following reduction steps:
\begin{itemize}
    \item Calibration files were created using the `Dark' and `Wavelength Solution 2D' recipes, applied to the dark frame and the Argon lamp calibration snapshot taken as part of the observations;
    \item A bad pixel map was created combining the results of the `Hot Bad Pixel Map' and `Cold Bad Pixel Map' recipes, which have been applied   respectively to a set of 15 dark frames and a set of 5 daytime Wollaston disperser flat frames for each filter (Y, J, H, K1, and K2). The calibration files were chosen from the Gemini Data Archive to be the closest in time to the observations;
    \item The data were reduced applying the `Calibrated Datacube Extraction' recipe, using the above-mentioned newly created calibration files. This recipe also includes an automatic search and characterisation of the four satellite spots, storing in the header the location and peak flux (in ADU) of all the spots, for each wavelength channel;
    \item The flux-calibrated cubes were oriented using the internal GPI recipe `Rotate North Up'.
\end{itemize}


\begin{table*}[ht]
\centering
\caption{Astrometry and photometry of the companion candidate for all three
datasets}
\begin{tabular}{lcccccc}
\hline \hline
\multicolumn{1}{c|}{Date of obs.}       & FPF                  & Separation        & P.A.                 & Projected semi-major axis              & Abs. Mag.                   \\
\multicolumn{1}{c|}{}     & $5\sigma$            & {[}arcsec{]}      & {[}deg{]}                      & {[}au{]}       & {[}mag{]}             \\ \hline
\multicolumn{1}{c|}{30/05/2016} & $4.4 \times 10^{-4}$ & $0.180\pm0.014$ & $152.35\pm4.46$ & $12.30\pm0.97$ & $M_{L'}=6.12\pm0.14$ \\
\multicolumn{1}{c|}{21/06/2018} & $3.6 \times 10^{-5}$ & $0.167\pm0.014$ & $170.27\pm4.81$ & $11.42\pm0.97$ & $M_{L'}=6.28\pm0.11$       \\
\multicolumn{1}{c|}{12/08/2018} & $1.00\times10^{-13}$ & $0.155\pm0.012$ & $176.90\pm3.71$ & $10.60\pm0.83$ & $M_{H}=6.89\pm0.06$       \\
\hline                 
\label{Tab: astrophoto}
\end{tabular}
\tablefoot{Given the small angular separation of the companion, the false probability fraction (FPF) values were evaluated on the classically ADI reduced images following the prescription in \citet{snr_mawet2014}, which accounts for small sample statistics. The final magnitudes are absolute values calculated taking into account the distance to the target and its uncertainties.}
\end{table*}

\begin{figure}[!]
        \centering
                \includegraphics[width=\columnwidth]{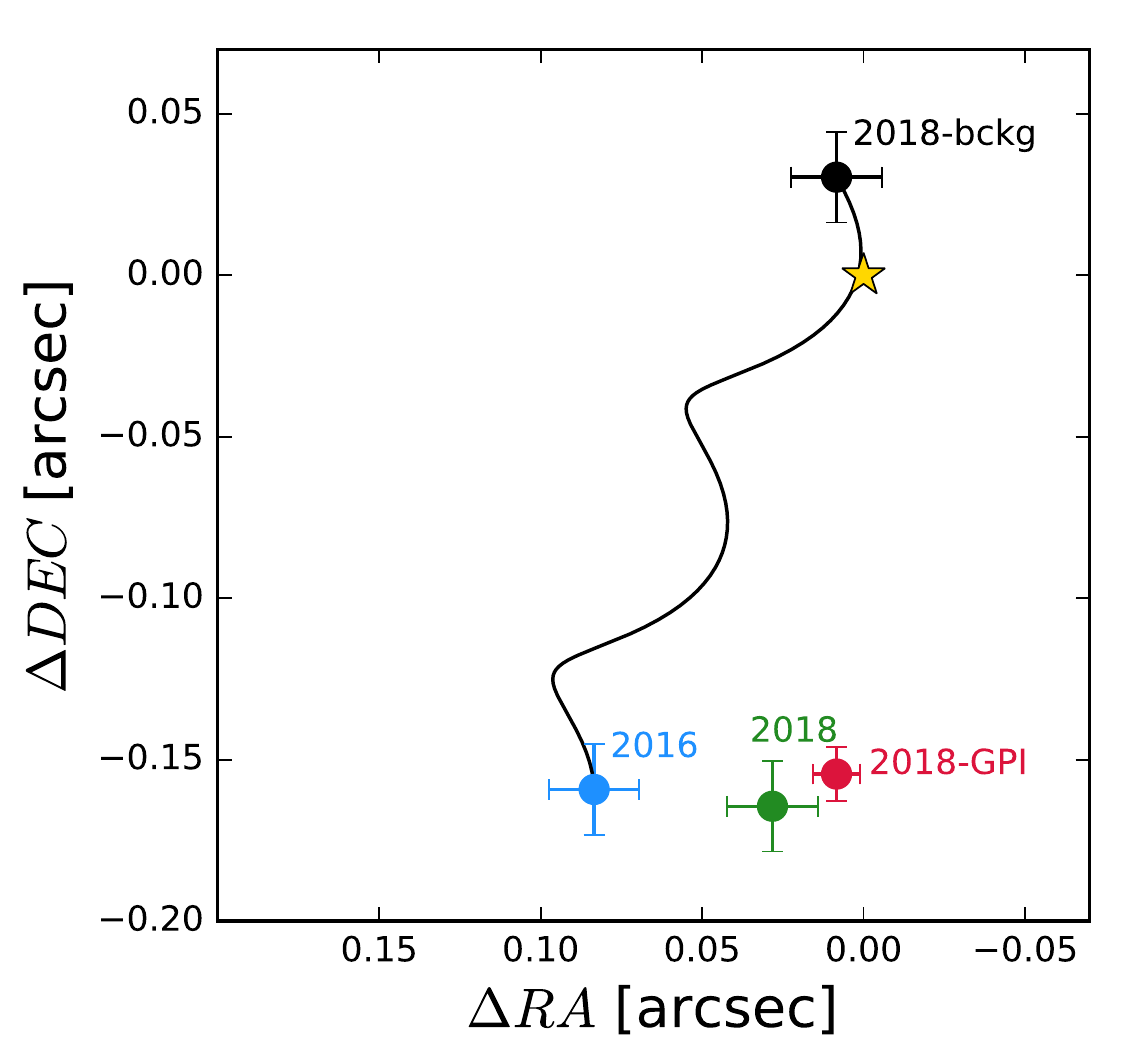}
                \caption{Proper motion analysis of the companion showing the astrometry for the three epochs. The black data point is the position that the companion would have at the epoch of the GPI observation if it were a background star with no motion, using its position in 2016 as starting point and considering the proper motion of the host star. The companion is clearly co-moving with the star (shown in yellow).} 
        \label{Fig: propermotion}\
\end{figure}

\section{Analysis and results}
The final classically ADI reduced images for all the three epochs are shown in \autoref{Fig: detection}. A close-in companion is clearly visible in all three epochs south of the star.
\subsection{Astrometry and photometry}  
\begin{figure}[t!]
        \centering
                \includegraphics[width=\columnwidth]{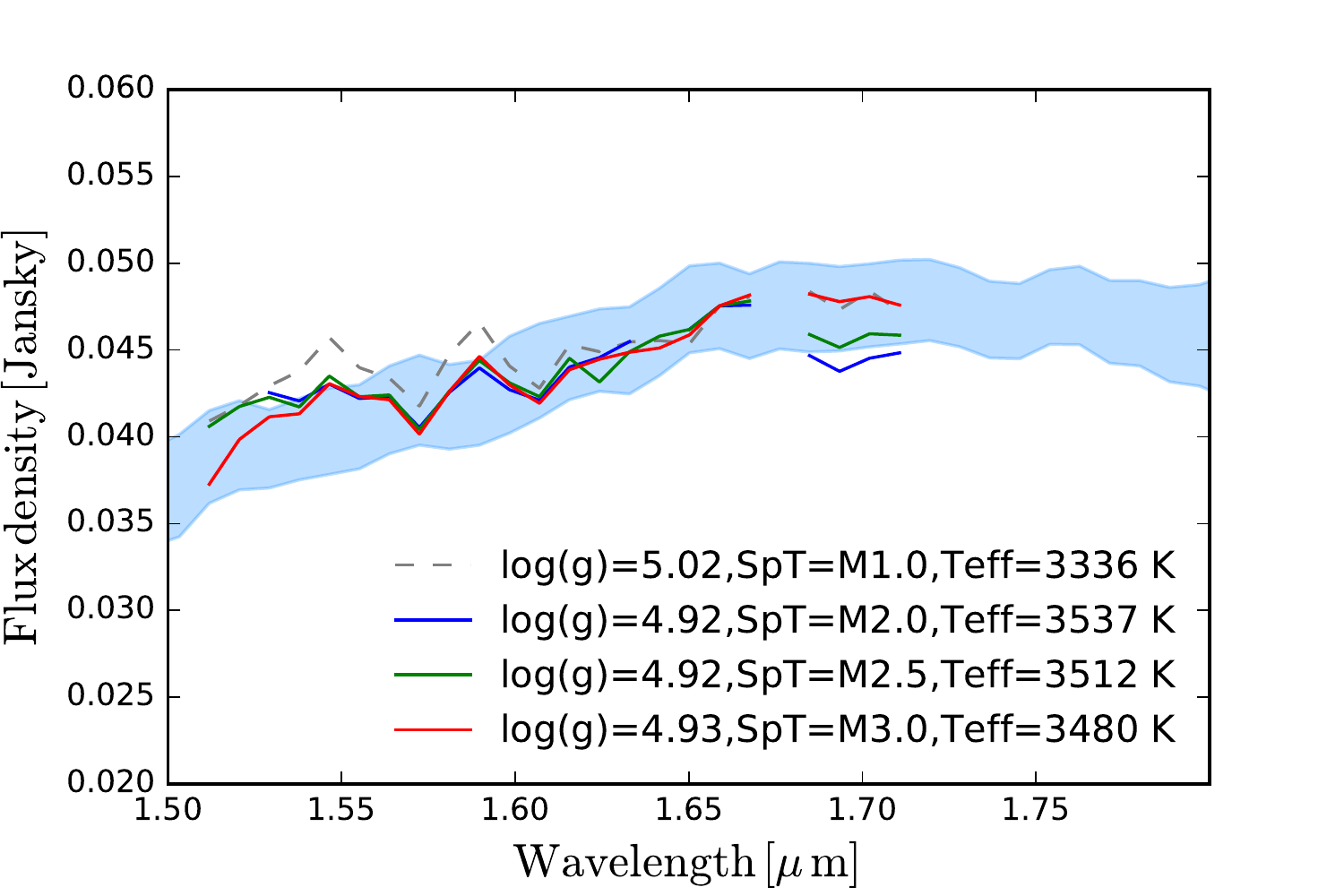}
                \caption{Comparison between the spectrum of the companion and observed spectra of early M dwarfs. The blue shaded area is the flux density of the companion in the GPI H-band, in Jansky. The spectrum is the weighted average of the spectra extracted from the 76 GPI datacubes and the area encompass the uncertainties (derived from the uncertainty on the flux of the host star). The solid lines are three spectra from the CARMENES stellar spectral library, for various T$\mathrm{_{eff}}$ and $\mathrm{log\,g}$ values (evaluated in \citealt{Hintz_2019}) and the dotted grey line is an additional spectrum of an M1 object.}
        \label{Fig: spectra}\
\end{figure}

To analyse the two NaCo datasets we used the $ANDROMEDA$ \mbox{\citep{Cantalloube_2015}}\mbox{\footnote{\url{http://www.theses.fr/2016GREAY017}}} package, which  uses a maximum likelihood estimation approach together with negative fake signal injection to evaluate the astrometry and photometry of a companion in an ADI dataset.
The algorithm needs as inputs the reduced frames (corrected for the AGPM throughput), the parallactic angles, and an unsaturated and exposure time-scaled image of the central star. Since we were interested in analysing only the known companion, we set the Inner Working Angle and Outer Working Angle keywords to $0.2\,\lambda/D$ and $20\,\lambda/D$, respectively (we refer to \mbox{\citealt{Cantalloube_2015}} for a detailed explanation of the $ANDROMEDA$ package). The final x and y offsets 
(and relative $3$$\sigma$ uncertainties) were converted into separation and position angle using a platescale for NaCo of $27.2$ mas/pix, assuming a conservative error of 0.5 pixels on the centring of the frames, and correcting for the true north offset of $0\fdg486\pm0\fdg180$ (Launhardt et al., in prep.). Given the target's distance and $L'$ band magnitude (see \mbox{\autoref{table: basic}}), we converted the flux evaluated with $ANDROMEDA$, and relative $3$$\sigma$ uncertainties, into an absolute $L'$ magnitude for both epochs accounting for the uncertainties on the host star magnitude and distance from the system.
The final astrometry and photometry values for the two NaCo epochs, as well as the GPI epoch, are given in \mbox{\autoref{Tab: astrophoto}}.\\
\indent For the GPI dataset we evaluated astrometry and photometry of the companion in a slightly different way since no unsaturated exposure of the central star was obtained.\\
For the astrometry, we made use of the satellite spots (visible in all the reduced frames) to create a PSF reference: we first averaged the two satellite spots in each frame, and then we averaged over the 76 frames, obtaining a PSF for each spectral channel. We use this PSF, together with the $ANDROMEDA$ package, to obtain the astrometry of the companion (as was done for the NaCo datasets) in each spectral cube. The final astrometry is the weighted mean of the astrometric positions at each wavelength, and is given in \mbox{\autoref{Tab: astrophoto}} taking into account the GPI pixel scale of $14.166$ mas/pix, the additional true north offset of $0.10\pm0.13^{\circ}$ as reported in \citealt{Rosa_2015}, and a conservative error on the centring of 0.5 pixels.\\
To obtain the photometry of the companion we calibrated the cubes extracted in Section 3.2 in the following way:\\
\begin{itemize}
    \item For each spectral channel, we averaged the satellite spots peak flux (stored in the header), obtaining a mean satellite flux in ADU, and relative standard deviation;
    \item We then converted the frame from ADU to physical units, using the following equation (as detailed on the GPI website\mbox{\footnote{\url{http://docs.planetimager.org/pipeline/usage/tutorial_spectrophotometry.html}}}):\vspace{8pt}\\
    \vspace{8pt}
    $\mathrm{frame[units]} = \frac{\mathrm{frame[ADU]}}{\mathrm{Satellite\,spectrum [ADU]}} \times \frac{\mathrm{Star\,Spectrum\,[units]}}{\mathrm{Star-to-Satellite\,Flux\,ratio}}$\\
    The `Star-to-Satellite Flux ratio' was calibrated by the GPI team\footnote{See footnote 4.}, and it is $=(2\times10^{-4})^{-1}$.
    The `Star Spectrum' (in the desired flux units) is obtained from the stellar spectrum fitted in Section 2. We accounted for the uncertainty on the `Star-to-Satellite Flux ratio',  the uncertainties on the stellar spectrum, and the standard deviation of the satellite spots flux;
    \item To account for possible contamination from the stellar halo, we median combined all the frames in each spectral channel, and then subtracted this median from each photometrically calibrated cube; 
    \item We then extracted a spectrum for the companion from each
    median-subtracted, photometrically calibrated cube, fitting a Gaussian to the companion to get the peak flux. The final spectrum is the weighted average of the spectra in all cubes.
\end{itemize}
The final spectrum of the companion is shown in \autoref{Fig: spectra}. We integrated this spectrum over the  NaCo H-band filter, obtaining a NaCo H-band apparent magnitude of $11.07\pm0.06$. This corresponds to an absolute magnitude of $6.89\pm0.06 $. The final astrometry and photometry for the companion is given in \autoref{Tab: astrophoto}.\\
The close separation makes it unlikely for the companion to be a background star. Nevertheless, we evaluated the position that the object would have on the sky at epoch 2018, starting from its position in epoch 2016, if it were a background object with no significant proper motion. The results are shown in \autoref{Fig: propermotion}. The object is clearly co-moving with the host star, at a projected separation of $\sim$11 au.

\subsection{Physical properties}
\begin{figure}[!]
        \centering
                \includegraphics[width=\columnwidth]{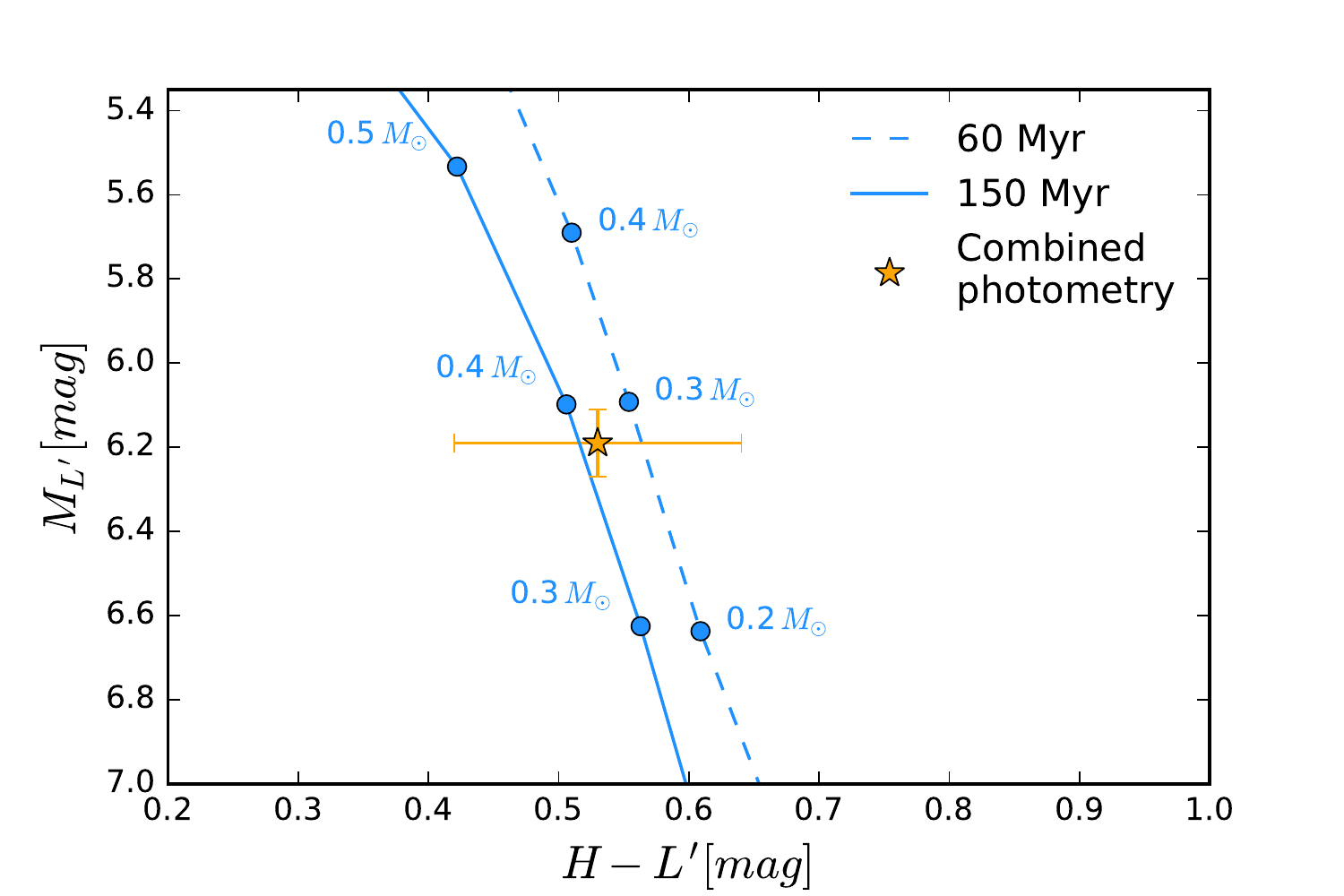}
                \caption{Colour-magnitude diagram showing the weighted mean $L'$-band magnitude derived from the 2016 and 2018 NaCo datasets, together with the $H$-band magnitude derived from the GPI dataset. We plot the evolutionary tracks for the BT-Settl models from \mbox{\citet{Allard2012}}, for ages of 60 and 150 Myr. The photometry does not allow us to distinguish between the two age estimates.}
        \label{Fig: mass_estim}\
\end{figure}

We compared the GPI H-band spectrum with observed spectra of early M dwarfs from the stellar spectral library \mbox{\footnote{\url{http://carmenes.cab.inta-csic.es/gto/jsp/reinersetal2018.jsp}}} of the CARMENES survey \citep{Reiners_2018}, which is the first large library of M dwarfs with high-resolution spectra in the infrared.
We plot three of the best matching spectra (binned to the GPI H-band resolution) in \autoref{Fig: spectra}, a non-matching spectrum (dotted grey line) for comparison, and the H-band spectrum of HD\,193571 B. From the comparison, 
we can infer  a surface gravity of $\mathrm{log\,g}\sim4.9$, a temperature of $\sim3500$ K, and 
a spectral type between M3 and M2, which seem to fit the data reasonably well. However, a high-resolution and/or broader band spectrum would be needed to properly constrain the surface gravity and spectral type of the companion.\\
We estimated the mass of the companion using the BT-Settl evolutionary tracks \citep{Allard2012}\footnote{\url{http://svo2.cab.inta-csic.es/theory/newov/}}, by comparing them with the observed $L'$- and $H$-band photometry. In the colour-magnitude diagram of \autoref{Fig: mass_estim} we show the companion $L'$-band absolute photometry of $6.19\pm0.08$ mag (evaluated as the weighted mean of the two NaCo epochs), as well as evolutionary tracks for two representative ages of 60 Myr (dashed line) and 150 Myr (solid line). As shown in \autoref{Fig: mass_estim}, the photometry does not allow us to distinguish between the two age estimates, so we  use both age values in the rest of the paper. 
We interpolated the BT-Settl models to estimate the mass of the companion for both $L'$- and $H$-band photometry, in mass steps of 0.034 dex. Taking into account the photometric uncertainty in both bands, we obtained a weighted mass of $0.395\pm0.007\,M_{\odot}$ for an age of 161 Myr, and $0.305\pm0.025\,M_{\odot}$ for an age of 66 Myr.

\subsection{Orbital motion}  
The astrometry of the companion between the three epochs shows signs of orbital motion.
Following the prescription in \mbox{\citet{Pearce2015}}, we can explore the possible orbital solutions for a companion imaged over a short orbital arc, using the dimensionless parameter $B$  ($\sqrt{B} = V_{\rm sky} / V_{\rm esc}$ is the sky-plane velocity of the companion divided by the escape velocity), and the direction of motion $\varphi$, where $\varphi=0^{\circ}$ is motion along a vector from the primary to the companion.\\
We assumed a total system mass of $2.6\pm0.1\,M_{\odot}$ (for an age of 161 Myr) and $2.55\pm0.1\,M_{\odot}$ (for an age of 66 Myr) and we derived\footnote{\url{http://drgmk.com/imorbel/}} $B$ and $\varphi$ for the three epochs (NaCo\,2016, NaCo\,2018, and GPI\,2018). For both age estimates the values agree within the uncertainties, and we obtain $B=0.25^{+0.16}_{-0.11}$ and \mbox{$\varphi=100\pm15^{\circ}$}, which leads to a minimum semi-major axis of \mbox{$a_\mathrm{min}=8.20\pm1.77$ au}
(see eq.\,(5) in \citealt{Pearce2015}).
Following \citet{Pearce2015}, we can draw the following conclusions:
\begin{itemize}
    \item Even considering the uncertainties, the $B$ value is $<1$, so the companion's sky-plane motion is below the escape velocity. While the object could be unbound if the line of sight velocity (or separation) is high, this is unlikely;
    \item We cannot place constraints on the eccentricity of the orbit, meaning that a circular orbit cannot be ruled out (this will have an impact on our stirring mechanisms study in Section 4);
    \item We can place a loose upper limit of $\sim$$80\,^{\circ}$ on the inclination.
\end{itemize}  
We also explored the possible orbital motion parameters using the python package $orbitize$\footnote{\url{https://orbitize.readthedocs.io/en/latest/}} with the Orbit For The Impatient (OFTI) algorithm detailed in \mbox{\citet{Blunt_2017_OFTI}} (see Appendix C). 
While the uncertainties on the astrometry and the limited amount of datapoints do not   place any meaningful constraints on the orbital elements, the periastron distance is restricted to $\lesssim$15 au. This result is confirmed by exploring the possible orbital parameters using the method of \mbox{\citet{Pearce2015}}. Therefore, if the companion's orbit is nearly coplanar with the disc, the entire orbit should be interior to the disc, otherwise the companion would have disrupted the disc on a dynamical timescale.
Assuming a circular orbit and a semi-major axis of $11$ au, the companion would have a minimum period of $\sim$23 years, implying that a baseline of several years would be needed before any additional astrometric datapoint could provide better constraints on the orbital elements.
The companion is  massive enough that even in the unlucky case of an almost face-on orbit (i$\,\sim$$\,1^{\circ}$) 
it would produce a radial velocity signal strong enough to be detected (semi-amplitude $K\gtrsim120$ m/s); however, this would also require a time baseline of many years.


\section{Stirring mechanisms}
The relative importance of self- and companion-stirring mechanisms is a non-trivial problem. It depends on the companion's physical and orbital parameters, the host star age and mass, and the disc mass in solids.
The equations used in this section are from \citet{Wyatt_2008} and \citet{MustillWyatt2009}, and are summarised in Appendix B. We note that to be consistent with the underlying assumptions of these two  papers, we use the black-body disc radius of 62 au while working with equations from \citet{Wyatt_2008}, and the corrected disc radius of 120 au for the \citet{MustillWyatt2009} equations (see Appendix B). That is, the model in \citet{Wyatt_2008} uses parameters derived by fitting to black-body radii, while the model of \citet{MustillWyatt2009} uses orbital dynamics, so is based on physical disc radii.\\
\begin{figure*}[t!]
    \centering
        \includegraphics[width=400pt]{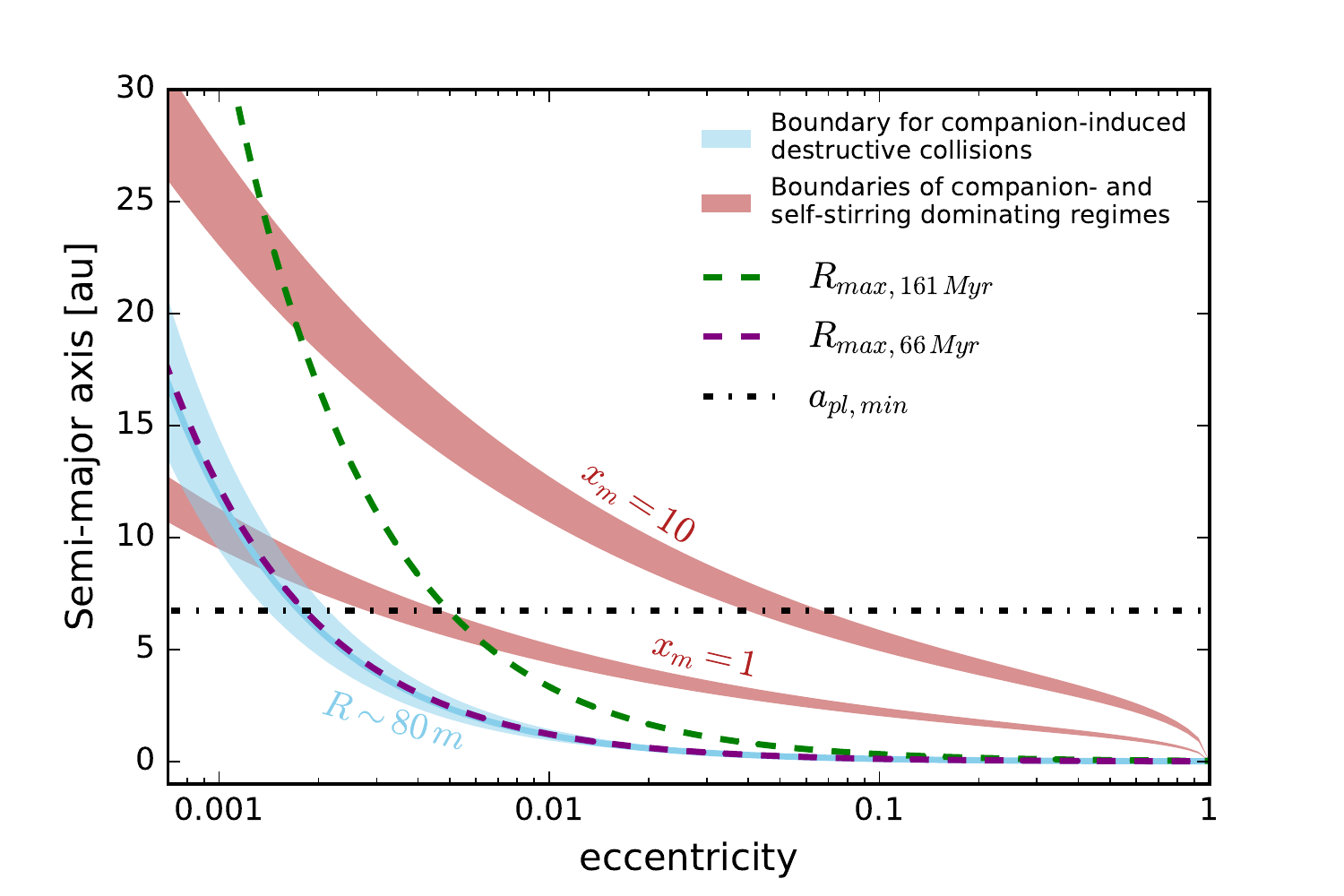}
        \caption{Boundaries between a self-stirring and companion-stirring dominated disc. The light blue lines mark the ($a_\mathrm{pl}$,$e_\mathrm{pl}$) parameter space in which the companion would be able to stir planetesimals of size R to destruction velocities at a distance of $120$ au. The shaded area around the solid light blue ($R=80$ m) line takes into account the errors on  the disc size and the stellar mass. The dashed purple line shows the $R_\mathrm{max}$ for 66 Myr (close to the solid light blue line) and the dashed green line shows the $R_\mathrm{max}$ value for the 161 Myr case. The shaded red areas indicate the boundaries between the self-stirring and companion-stirring dominated cases, for a fixed distance and companion mass, and for two representative $x_\mathrm{m}$ values; accounting for errors on disc size, stellar mass, and companion mass (the areas encompass both age estimates). The horizontal dotted black line is the lowermost boundary of the minimum possible companion semi-major axis calculated in Section 3.3. The companion dominates the stirring process only for combinations of $a_\mathrm{pl}$ and $e_\mathrm{pl}$ lying above  the light blue curve (the companion can stir planetesimals at the disc distance) and the red curve (the companion stirs the disc faster than the disc stirs itself).}
        \label{Fig: stirring}\
\end{figure*}
Assuming that the mutual inclination between the plane of the orbit and the disc is not too large, there are two conditions that need to be satisfied for a companion to dominate the stirring process at a certain distance from the star, and at a given time: a) the companion must be able to stir planetesimals, at that location, to relative destructive velocities and b) the timescale for companion-stirring at that distance must be greater  than the self-stirring timescale.\\
\indent The first condition is encapsulated by Eqs (2) and (3) in Appendix B, which give the maximum distance at which a companion with a given semi-major axis $a_\mathrm{pl}$ and eccentricity $e_\mathrm{pl}$ can stir planetesimals above the disruption threshold velocity $v_\mathrm{rel}$.
This velocity is a function of the planetesimal size $R$ and, as shown by eq. (2), has a minimum at $R$$\sim$$80\,$m. We set this maximum distance equal to the estimated true disc radius of $120$ au,
 and we plotted the $a_\mathrm{pl}$-$e_\mathrm{pl}$ relationship in \autoref{Fig: stirring} for the $R=80$ m case (solid light blue curve). The companion would not be able to stir planetesimals at that distance if its semi-major axis and eccentricity were below this curve.
The planetesimals might be smaller or larger than $80$ m, and this would increase $v_\mathrm{rel}$ and push the light blue curve rightwards and upwards. While $R$ has a definite minimum (particles smaller than a certain size, typically around few $\mathrm{\muup m}$, would be blown away by radiation pressure from the central star) it is not straightforward to define a maximum $R$ value. We proceeded as follows: 
\begin{itemize}
    \item At any given time, there is a  maximum  size of planetesimals that  participate in the collisional cascade (because larger objects will have collision timescales longer than the stellar age). This maximum size $R_\mathrm{max}$ can be evaluated by inverting eq. (1). For a disc size of $62$ au, and with a fractional luminosity of the disc $f$, stellar mass and stellar luminosity as in \autoref{table: basic}, we have $R_\mathrm{max}=132$ m. This is the maximum value for $R$,
    assuming that the disc has been stirred for all of its life ($t_\mathrm{stir}=t_\mathrm{age}=66$ Myr. In the 161 Myr case we obtain $R_\mathrm{max}=790$ m);
    \item An internal perturber can influence the timescale of orbit crossings for planetesimals, and thus $t_\mathrm{stir}$ might be less than the stellar age (i.e. the disc was stirred more recently). We use eq. (4) to calculate this orbit crossing timescale $t_\mathrm{cross}$ as a function of the perturber properties (eccentricity, semi-major axis, and mass);
    \item We now have a revised value for the total time the disc has been stirred as $t_\mathrm{stir}=t_\mathrm{age}-t_\mathrm{cross}$, and consequently a revised $R_\mathrm{max}$ value as a function of the perturber properties (i.e. we have a relationship between $R_\mathrm{max}$, $a_\mathrm{pl}$, and $e_\mathrm{pl}$);
    \item Combining this relationship with Eqs. (2) and (3), we can trace $R_\mathrm{max}$ in the ($a_\mathrm{pl}$,$e_\mathrm{pl}$) parameter space.
\end{itemize}
As can be seen in \mbox{\autoref{Fig: stirring}}, when we plot this for the 66 Myr case, $R_\mathrm{max}$ is relatively small ($\sim$132 m along the curve) and almost overlaps with the $R=80$ m case. The $R_\mathrm{max}$ in the 161 Myr case is plotted with a dashed grey curve. The companion can stir the disc over most of the shown parameter space.\\
\indent The second condition requires that, at a given time and distance, the companion-stirring timescale is shorter than the self-stirring timescale. \mbox{\citet{MustillWyatt2009}} made such a study and defined the parameter $\Phi$ as the distance at which self and companion-stirring times are equal (see Appendix B). It is a function of the companion's properties (mass $m_{pl}$, semi-major axis $a_\mathrm{pl}$, and eccentricity $e_\mathrm{pl}$), the central star's mass, 
and the disc's mass in solids (expressed by the dimensionless parameter $x_\mathrm{m}$, see Appendix B). Since we are interested in  which stirring process is dominant at the location of the debris belt, we set $\Phi=120$ au and obtain the equilibrium relationship between self- and planetary-stirring. Tracing this line in the $(a_\mathrm{pl}$, $e_\mathrm{pl})$ parameter space marks the boundary between the domination of the two stirring processes, thus allowing us to investigate the combination of $a_\mathrm{pl}$ and $e_\mathrm{pl}$ for which the disc is dominated by self-stirring. Since there is a dependence on the $x_\mathrm{m}$ value as well, in
\autoref{Fig: stirring} we plotted two representative values for $x_\mathrm{m}$ of $1$ and $10$ (solid red lines). The curve for $x_\mathrm{m}=10$ lies above the $x_\mathrm{m}=1$ case because a more massive disc forms large planetesimals more quickly, and can thus self-stir earlier.
As discussed in \citet{MustillWyatt2009}, $x_\mathrm{m}\gtrsim10$ discs may be problematic as their high masses imply gravitationally unstable discs at earlier times when the gas was present. Thus, it is likely that the $x_\mathrm{m}=10$ line in \autoref{Fig: stirring} represents an upper limit to where the disc could be self-stirred.
Given an $x_\mathrm{m}$ value and fixing the companion mass to $0.25\,M_{\odot}$, any combination of eccentricity and semi-major axis above the curve would imply that companion-stirring is quicker than self-stirring at the distance of the disc, hence the companion-stirring would dominate the stirring process.
An additional constraint can be placed on the minimum semi-major axis, as discussed in Section 3.3, which is shown by the dashed black line in \autoref{Fig: stirring}.\\
\indent It is important to note that both conditions must be satisfied for the companion to dominate the stirring process, and this is true only for certain combinations of eccentricity and semi-major axis. In the plot it is clear how, given an eccentricity $\gtrsim0.1$, any semi-major axis places the companion above both curves, and thus the companion would   dominate. For eccentricities $\gtrsim0.002$, any $a_\mathrm{pl}$ would lie above the light blue curves (both for the $R$$\sim$80 m and for the $R_\mathrm{max}$ case), but only certain $a_\mathrm{pl}$ would satisfy the criterion for companion-induced stirring dominating over self-stirring (depending on the $x_\mathrm{m}$ value), so low-eccentricity companions must be closer to the disc to dominate the stirring. Finally, for extremely low eccentricities ($\lesssim0.002$) and small semi-major axes, the companion would not be able to stir planetesimals at the distance of the disc (below the light blue curve), and in any case the self-stirring would be dominant at that distance (below the red curve).\\
As shown in \autoref{Fig: stirring}, it is most likely that the companion is dominating the stirring process, and
self-stirring is relevant only when the companion has a very low  eccentricity (in combination with a small semi-major axis).


\section{Conclusions}
We presented the first detection of a close low-mass stellar companion around the A0 star HD\,193571. The three epochs obtained with VLT/NaCo and GPI confirm that the companion is co-moving with the host star, showing the potential of multi-band/multi-instrument follow-up to confirm direct imaging candidates.
Comparing $M_\mathrm{H}$ and $M_\mathrm{L'}$ band photometry to evolutionary tracks suggests a mass of $\sim$$0.305\pm0.025\,M_{\odot}$ for an age of 66 Myr ($\sim$$0.395\pm0.007\,M_{\odot}$ for the 161 Myr case), which would make it an M2-2.5 dwarf. Comparison to observed spectra seems to suggest a surface gravity of $\sim4.9$ and a temperature of $\sim3500$ K.
The orbital motion detected in the three epochs is not enough to place solid constraints on the orbital parameters, but allows us to confirm the co-motion with the host star and to exclude an edge-on orbit.\\
Given the projected separation of $\sim$11 au and a maximum periastron of $\sim$15 au, the companion appears to orbit interior to the circumstellar debris belt (inferred via SED IR-excess to be at $\sim$120 au). We investigated the plausibility that both self- and companion-stirring mechanisms are responsible for the currently observed debris belt radius. Since no constraints can be put on the eccentricity, we cannot exclude a fully self-stirring scenario for the disc. However, a small deviation from a circular orbit would result in the disc being dominated by companion-stirring (as shown in \autoref{Fig: stirring}) and if the orbit is sufficiently eccentric the disc will appear eccentric as well. The companion is likely responsible for the stirring of a disc that appears to be an order of magnitude further away, showing how a massive companion can influence a debris disc at large distances.\\
At the moment, only a handful of systems are suited for a study of stirring mechanisms, and the HD\,193571 system represents an important addition, containing the third known M-dwarf companion to a young star discovered to be orbiting within the primary's circumstellar disc, and the first one found around an A0-type star. \
In the future, radial velocity observations as well as a resolved image of the disc could be useful in deepening our understanding of this system.

\begin{acknowledgements}
A.M.B. thanks Christian Ginsky for the help in reducing the SPHERE/IRDIS data and the useful discussion afterwards, and Wolfgang Brandner for the help in reducing the GPI data.
A.M.B. would also like to thank the anonymous referee for the useful and constructive comments.
G.M.K. is supported by the Royal Society as a Royal Society University Research Fellow.
G.C. and S.P.Q. thank the Swiss National Science Foundation for the financial support under the grant number 200021$\_$169131.
J.O. and N.G. acknowledge financial support from the ICM (Iniciativa Cient\'ifica Milenio) via the N\'ucleo Milenio de Formaci\'on Planetaria grant. J.O. acknowledges financial support from the Universidad de Valpara\'iso, and from Fondecyt (grant 1180395).
N.G. acknowledges support from project CONICYT-PFCHA / Doctorado Nacional / 2017 folio 21170650.
A.M. and A.Q. acknowledge the support of the DFG priority program SPP 1992 “Exploring the Diversity of Extrasolar Planets” (MU 4172/1-1 and QU 113/6-1). 
This research has made use of the Washington Double Star Catalog maintained at the U.S. Naval Observatory, and of data products from the Two Micron All Sky Survey, which is a joint project of the University of Massachusetts and the Infrared Processing and Analysis Center/California Institute of Technology, funded by the National Aeronautics and Space Administration and the National Science Foundation. This work has made use of data from the European Space Agency (ESA) mission {\it Gaia} (\url{https://www.cosmos.esa.int/gaia}), processed by the {\it Gaia} Data Processing and Analysis Consortium (DPAC, \url{https://www.cosmos.esa.int/web/gaia/dpac/consortium}). Funding for the DPAC has been provided by national institutions, in particular the institutions participating in the {\it Gaia} Multilateral Agreement.
This research made use of Astropy,\footnote{http://www.astropy.org} a community-developed core Python package for Astronomy \citep{Astropy_2013, Astropy_2018}.
\end{acknowledgements}
\bibliographystyle{aa}
\bibliography{Bibliograpy.bib}

\clearpage
\begin{appendix}

\section*{Appendix A - IRDIS disc non-detection}
We observed the target with SPHERE/IRDIS at the VLT in coronagraphic Differential Polarisation Imaging (DPI) mode on 26 September 2018, using the $H$ broad-band filter.\\
We took eight polarimetric cycles, each consisting of four data cubes, one per half wave plate (HWP) position. Each data cube consisted of four individual exposures with exposure times of 32 s.
The science observations were bracketed with 2-second exposures, to create an unsaturated PSF reference for the central star.\\
The data were reduced following the prescription in \mbox{\citet{Ginski_2016}}, obtaining the radial Stokes components $Q_{\Phi}$ and $U_{\Phi}$ \mbox{\citep[see][]{Schmid_2006}}, where $Q_{\Phi}$ would contain any polarisation signal coming from dust scattered light, and it is shown in \autoref{Fig: IRDIS}. No emission is visible at the expected location of the disc ($\sim$$1\farcs75$) or anywhere else. The faint emission from the centre is due to the stellar halo, and the spider is vaguely visible extending approximately in the  north-south direction.\\

\renewcommand\thefigure{B.\arabic{figure}} 
\setcounter{figure}{0}
\begin{figure}[t!]
        \centering
                \includegraphics[width=\columnwidth]{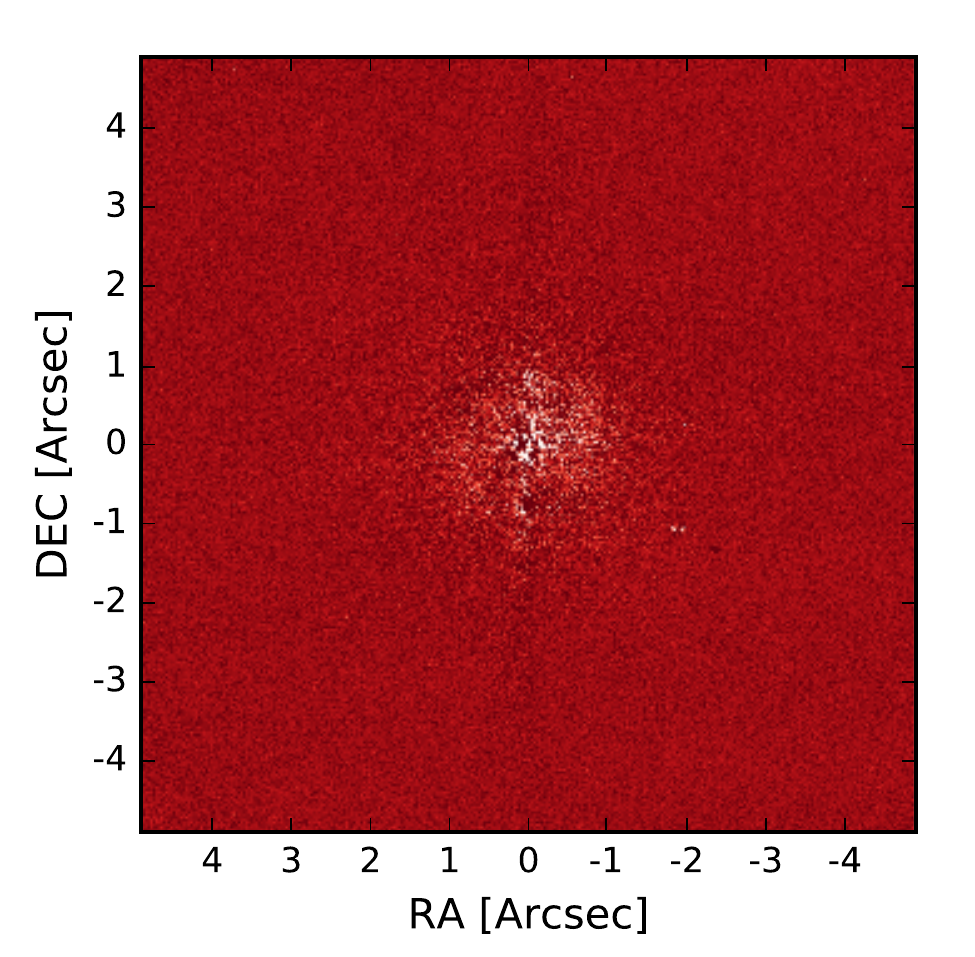}
                \caption{DPI data taken with SPHERE/IRDIS, with a total field of view of $\sim$$10"\times10"$, given a pixel scale for IRDIS of $12.25$ mas/pix. No polarised signal from the disc scattered light is visible. The image is oriented with north up and east left.}
        \label{Fig: IRDIS}\
\end{figure}

\section*{Appendix B - Stirring mechanisms}
\subsection*{Self-stirring}
From \citet{Wyatt_2008} the maximum fractional luminosity $f_\mathrm{max}$ of a planetesimal belt at distance $r$ around a star of mass $m_{\star}$, luminosity $L_{\star}$, and age $t_\mathrm{age}$ is 
\begin{equation}
    f_\mathrm{max}=0.58\times10^{-9}r^{7/3}\left ( dr/r \right )R^{0.5}_\mathrm{max}Q^{\star 5/6}_\mathrm{D}e^{-5/3}m^{-5/6}_{\star}L^{-0.5}_{\star}t^{-1}_\mathrm{age}
,\end{equation}
where $R_\mathrm{max}$ is the maximum size of the planetesimals that are participating in the cascade at that given time (called $D_\mathrm{c}$ in \citealt{Wyatt_2008}), $Q^{\star}_\mathrm{D}$ is the planetesimal strength in $\mathrm{Jkg^{-1}}$, $e$ is the mean planetesimal eccentricity, and $dr/r$ is the relative width of the planetesimal belt. It was found (see \citealt{Wyatt_2008}) that the population of debris discs around A stars can be fitted assuming $Q^{\star}_\mathrm{D}=150\,\mathrm{Jkg^{-1}}$, $e=0.05$, and $dr/r=0.5$. All of this assumes that the disc has been stirred for its whole lifetime (i.e. $t_\mathrm{stir}=t_\mathrm{age}$). The disc evolution model developed in \citealt{Wyatt_2008} is SED-based, and therefore the planetesimal belt distance $r$ refers to the black-body radius $R_{\mathrm{BB}}$ of 62 au, inferred via SED fitting.
\subsection*{Companion-stirring}
From \citet{MustillWyatt2009}, the threshold velocity above which collisions between planetesimal of size R become destructive is
\begin{equation}
    v^{\star}_\mathrm{rel}(R)=\left [ 0.8\left ( \frac{R}{80\,\mathrm{m}} \right )^{-0.33}+0.2\left ( \frac{R}{80\,\mathrm{m}} \right )^{1.2} \right ]^{0.83}\,\mathrm{ms^{-1}}
\end{equation}
A companion of mass $m_\mathrm{pl}$ internal to the disc on an orbit of semi-major axis $a_\mathrm{pl}$ and eccentricity $e_\mathrm{pl}$, around a primary of mass $m_{\star}$, would be able to stir planetesimals to catastrophic collisions only up to a maximum distance $a^{\star}$:
\begin{equation}
    a^{\star}(R)=3.8\mathrm{au}\left ( \frac{e_\mathrm{pl}}{0.1} \right )^{2/3} \left ( \frac{m_{\star}}{1\,M_{\odot}} \right )^{1/3}\left ( \frac{a_\mathrm{pl}}{1\mathrm{au}} \right )^{2/3}\left ( \frac{v^{\star}_\mathrm{rel}(R)}{1\mathrm{kms^{-1}}} \right )^{-2/3}
\end{equation}
In addition, it is possible to calculate the timescale for orbit crossing of planetesimals at a distance $a$ as
\begin{equation}
\begin{split}
    t_\mathrm{cross}\sim1.53\times10^{3}\frac{\left ( 1-e^{2}_\mathrm{pl} \right )^{3/2}}{e_\mathrm{pl}}\left ( \frac{a}{10\mathrm{au}} \right )^{9/2} \\
    \times \left ( \frac{m_{\star}}{M_{\odot}} \right )^{1/2}\left ( \frac{m_\mathrm{pl}}{M_{\odot}} \right )^{-1}\left ( \frac{a_\mathrm{pl}}{1\mathrm{au}} \right )^{-3}\mathrm{yr}
\end{split}
\end{equation}
\subsection*{Companion-stirring versus self-stirring}
\citet{MustillWyatt2009} also defined the parameter $\Phi$ as the distance boundary between self-stirring and companion-stirring at a fixed age as
\begin{equation}
    \begin{split}
        \Phi=630\,\mathrm{au}\left ( 1-e^{2}_\mathrm{pl} \right )^{-1}e^{2/3}_\mathrm{pl}\left ( \frac{m_\mathrm{pl}}{M_{\odot}} \right )^{2/3}\\\times\left ( \frac{a_\mathrm{pl}}{1au} \right )^{2}\left ( \frac{m_{\star}}{M_{\odot}} \right )^{-4/3}x^{-0.77}_\mathrm{m}
    \end{split}
,\end{equation}
where the dimensionless parameter $x_\mathrm{m}$ is a scaling factor relating the disc surface density to the minimum mass solar nebula density (see \mbox{\citealt{MustillWyatt2009}} and \mbox{\citealt{KenyonBromley2008})}.\\
The model developed in \citealt{MustillWyatt2009} is a dynamic model that depends on the physical structure of the disc, and therefore on the real disc size of 120 au (see Section 2).

\renewcommand\thefigure{C.\arabic{figure}} 
\setcounter{figure}{0}
\begin{figure*}[ht!]
        \centering
                \includegraphics[scale=0.45]{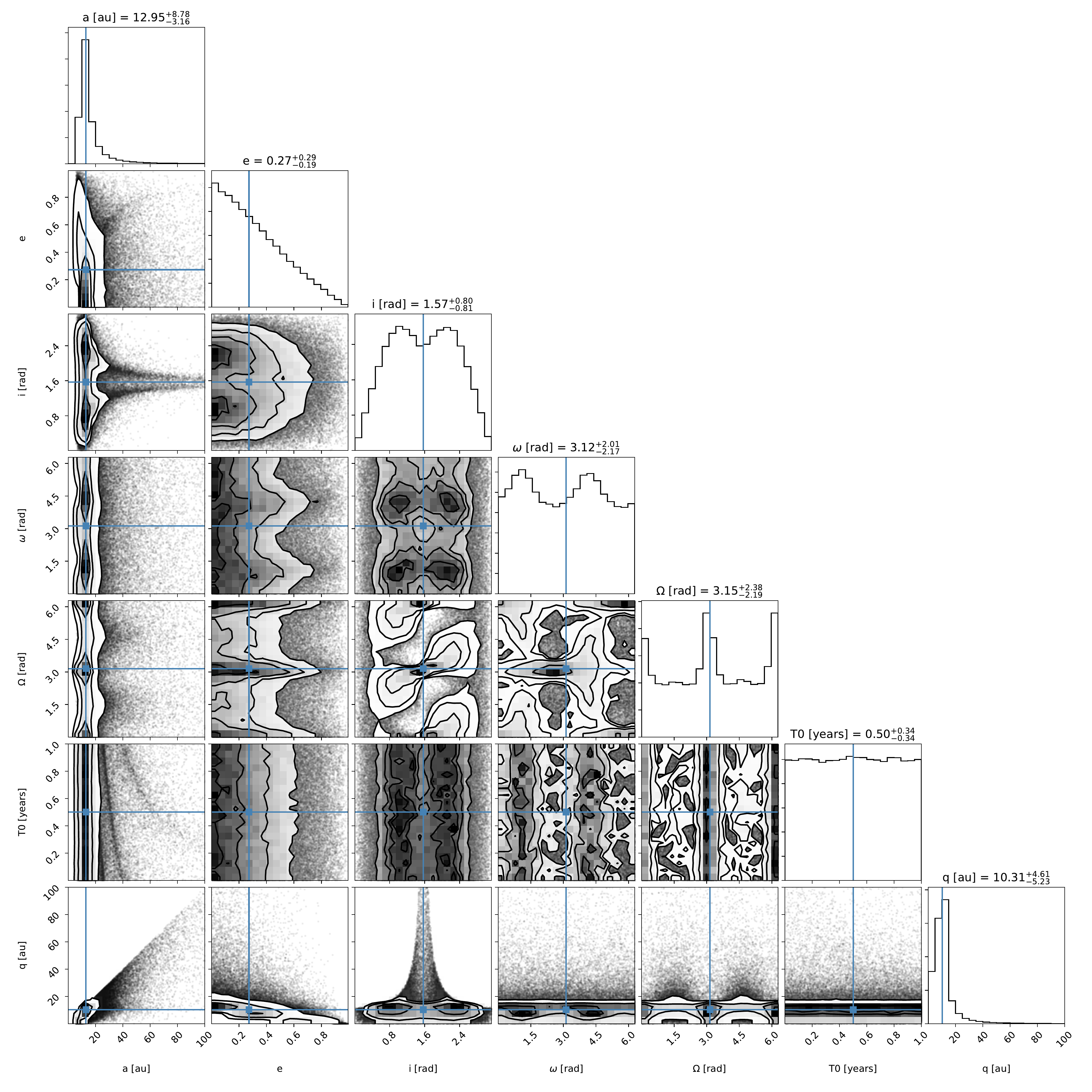}
                \caption{Posterior distribution function for the orbital parameters derived with the $orbitize$ package using the OFTI implementation.}
        \label{Fig: OFTI}\
\end{figure*}
\section*{Appendix C - Orbital constraints with OFTI}
We explored the possible orbital motion parameters using the python package $orbitize$ with the Orbit For The Impatient (OFTI) algorithm detailed in \mbox{\citet{Blunt_2017_OFTI}}. We used two total mass estimates: $2.6\pm0.1\,M_{\odot}$ (for an age of 161 Myr) and $2.5\pm0.1\,M_{\odot}$ (for an age of 66 Myr). 
We used a uniform prior for the semi-major axis, and in the epoch of periastron passage and argument of periastron. We used a $\mathrm{sin(i)}$ prior for the inclination angle, and a linearly descending prior for the eccentricity, with a slope of $-2.18$.
For both age estimates, the results agree within the error bar, and in \autoref{Fig: OFTI} we show the posterior distribution function for the 161 Myr case.
As shown in the figure, the uncertainties on the astrometry and the limited number of datapoints do not allow us to place any meaningful constraints on the orbital elements, but the periastron distance q is restricted to $\lesssim$15 au.

\end{appendix}


\end{document}